# Some reports of snowfall from fog during the UK winter of 2008/09

Curtis R. Wood[*] and R. Giles Harrison

Department of Meteorology, University of Reading, UK

**Abstract:** Snowfall during anticyclonic, non-frontal, and foggy conditions is surprising. Because it is often not forecast, it can present a hazard to transport and modify the surface albedo. In this report, we present some observations of snowfall during conditions of freezing fog in the UK during the winter of 2008/09.

## Introduction

Snow falling from freezing fog[†] during non-frontal conditions has been reported elsewhere (e.g. Parungo and Weickmann 1978, Koenig 1981, Van den Berg 2008). In those cases, it appears that anthropogenically-produced particles were present and contributed to the snow production, hence these events are usefully labelled Anthropogenic Snowfall Events (ASEs). Two further such snowfalls that occurred during non-frontal freezing fog conditions in the UK during the winter of 2008/09 are reported here.

## Event 1: 31 December 2008

Freezing fog was the sole weather reported at most synoptic stations across the UK and central Europe (radiosonde ascents showed a typical fog depth of 300–400 m) on 31 December 2008. Despite the absence of frontal activity (Figure 1) and cloud (Figure 2) over the UK, there were reports of snow at Hereford, Dudley, Gloucester, Watlington, Tadley, Eccles, and mentioned on the UK's BBC television weather forecast.

---

[*] c.r.wood @ reading.ac.uk
[†] We propose that since snow grains falling from freezing fog are drizzle-like, we term the precipitation as *swizzle* (a contraction of the words 'snow' and 'drizzle').

Some reports of snowfall from fog during the UK winter of 2008/09

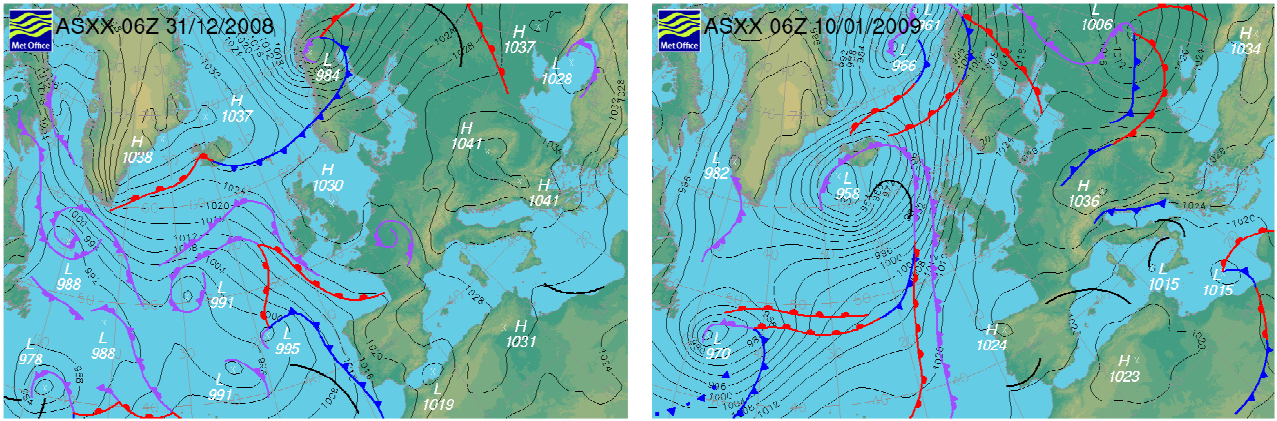

**Figure 1:** Synoptic analyses at 06 hours UTC, on 31 December 2008 (left) and 10 January 2009 (right). (With permission from the Met Office.)

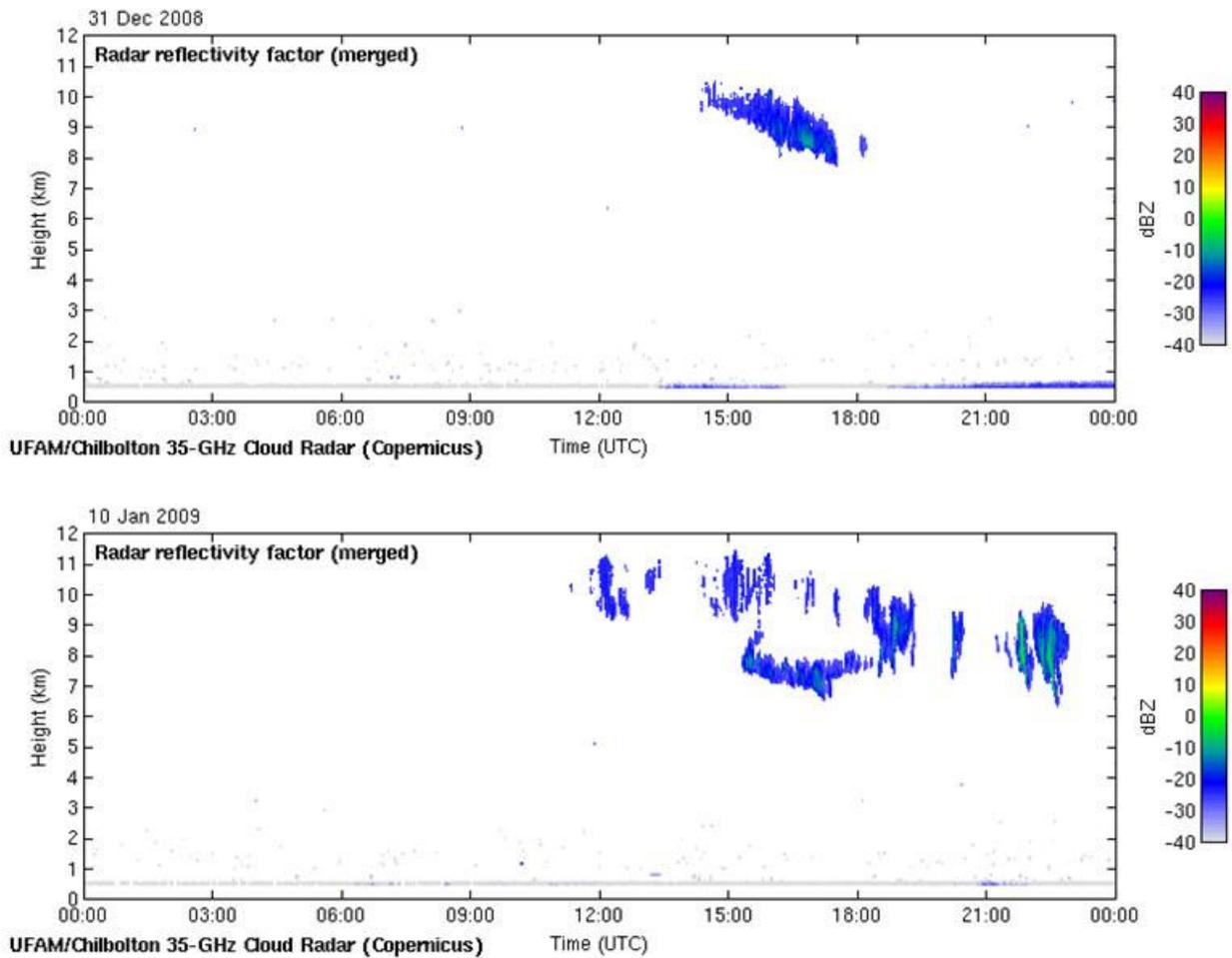

**Figure 2:** Cloud radar profiles for 31 December 2008 and 10 January 2009 at Chilbolton, Hampshire. In both cases no cloud was observed until after 12:00, and that was thin cirrus cloud above ~ 7 km. (With permission from CloudNet; www.cloud-net.org)

In Hereford, an opportunity presented itself to drive a car through and around Hereford to make observations of lying and falling snow, take photographs and record air temperature. The map (Figure 3) shows that most snow was observed within a kilometre of the city centre, and that towards the west there was snow observed several kilometres away in Breinton villages. The edges of the snow area were ascertained in all compass directions. Photos of falling and lying snow were taken at many of the locations (Figure 4), and the route and timings are listed in the Appendix.

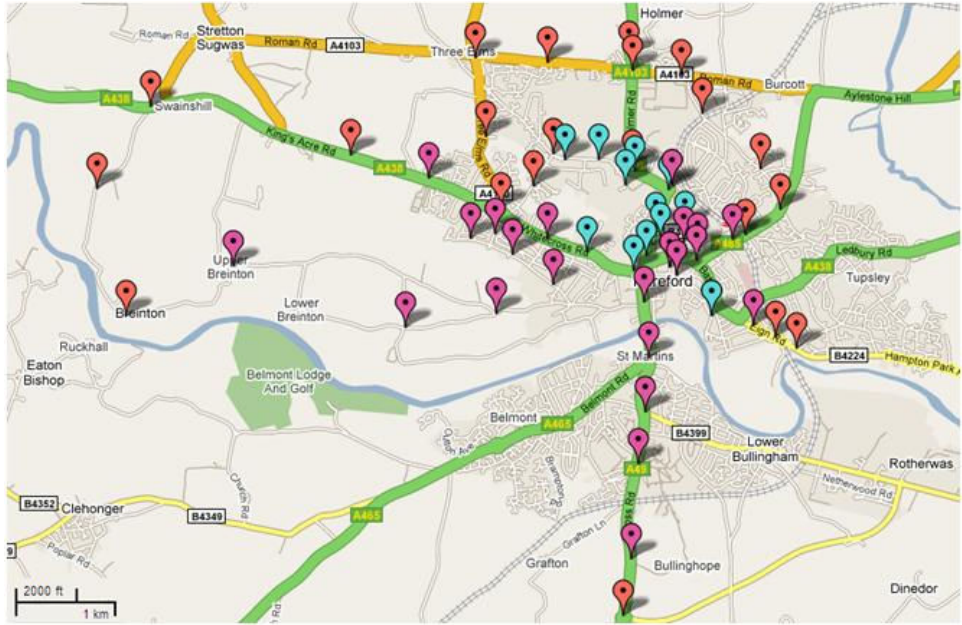

**Figure 3:** Central Hereford (map constructed using Google Maps), 31 December 2008 from 13:20–17:00. Blue pins are where falling snow was observed, purple pins are where lying snow was observed, and red pins are where there was no evidence of snow at all.

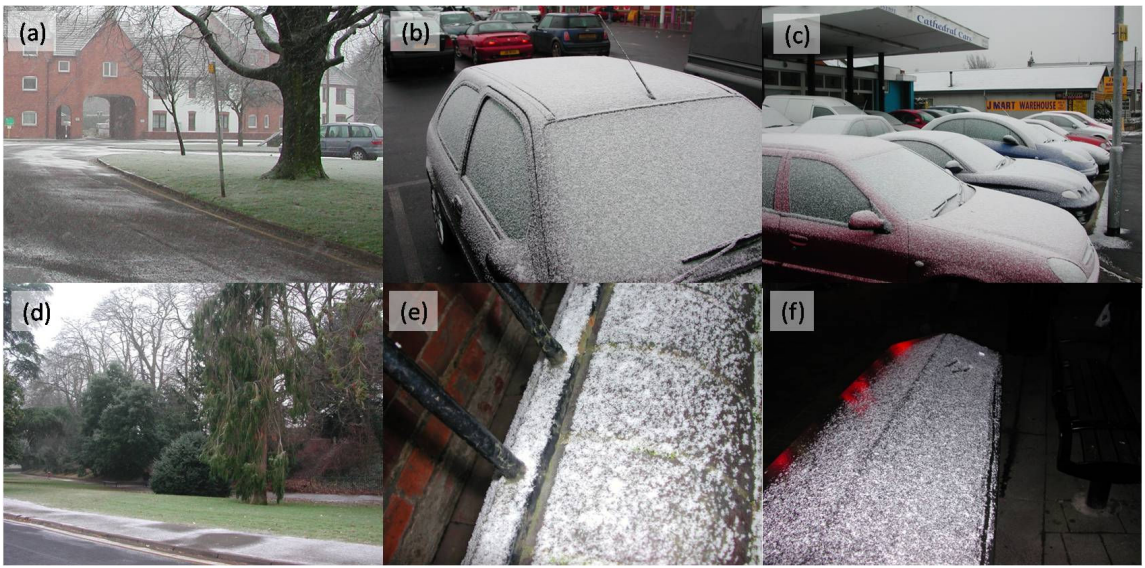

**Figure 4:** Photographs around Hereford on 31 December 2008 at times 13:34, 13:50 14:10, 14:25, 16:00, and 16:30 (the locations can be inferred from the Appendix). Note the falling snow in photograph (a).

In Gloucester, the maximum temperature was –1.2°C and the minimum was –5.9°C (Severn Tales Weather Station, Figure 5). Snow was seen falling at about 08:30 for 10 minutes in the Brooklands Park area of Longlevens (Bennett 2009), as marked on Figure 6.

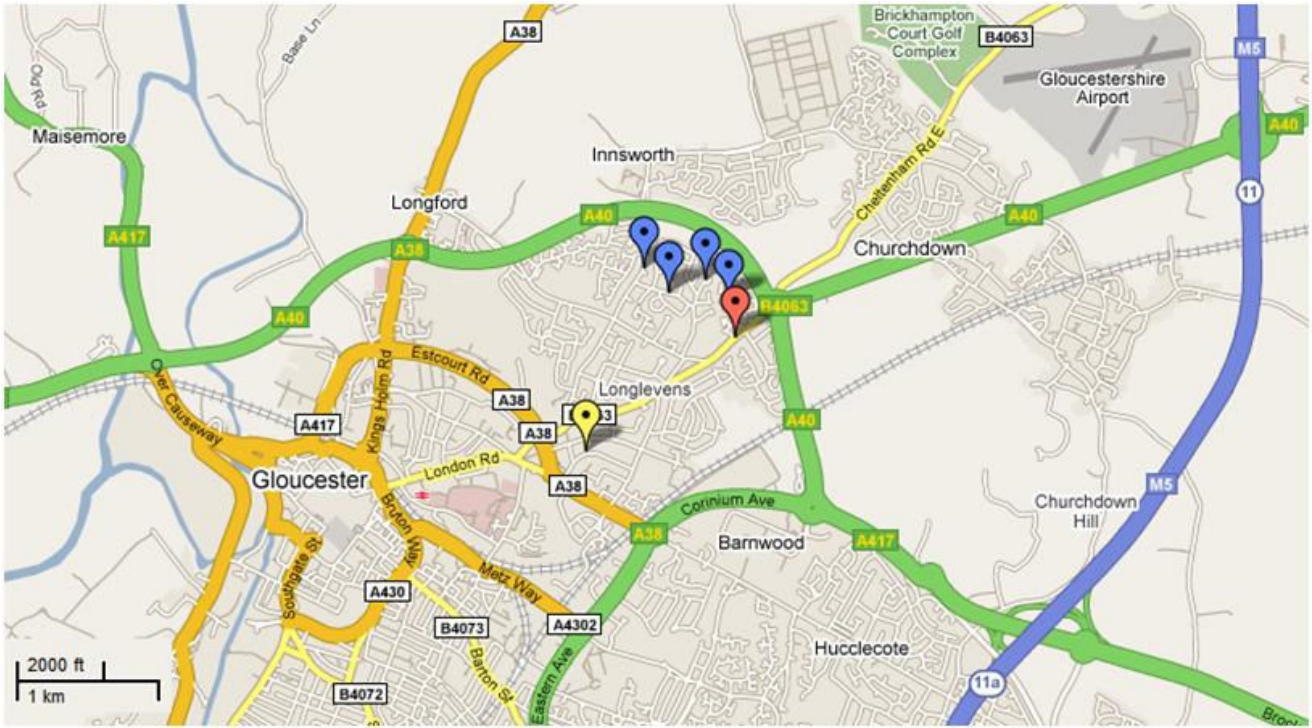

**Figure 5:** Gloucester (map constructed using Google Maps) on 31 December 2008. Blue pins are where snow was reported, the red pin is where there was no evidence of snow at all, and the yellow pin is SevernTales Weather Station.

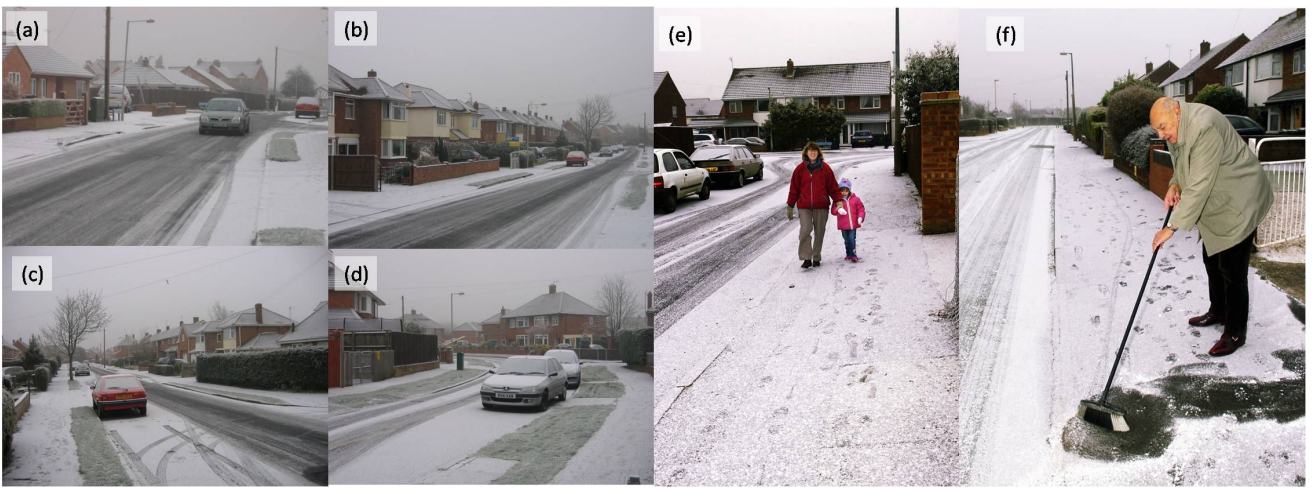

**Figure 6:** Photographs around the Longlevens area of Gloucester of the 31 December 2008 event of snowfall from fog. (With permission from Dave Bennett [a–d] and the Gloucester Citizen [e–f].)

## Event 2: 10 January 2009

Conditions were of freezing fog across southern UK (again in the absence of cloud (Figure 2) and frontal activity (Figure 1)), and temperatures until noon varied between –5.5 and –2.6°C at the University of Reading's Atmospheric Observatory (www.met.rdg.ac.uk/~fsdata/obshome.html). Following a general email around the Meteorology Department, responses were compiled to reveal the extent of falling and lying snow in and around Reading (Figure 7). There was sufficient snow to require car windscreens to be cleared and to remain lying all day. Almost all observers volunteered that they thought the snow was very slight, consisting of small flakes and 'like desiccated coconut'. A photograph was taken of the snow deposits on some garden furniture (Figure 8). Snowfall times varied slightly between observers, but most reports were at the times of 06:00, 09:00–10:00, and 11:00–12:00.

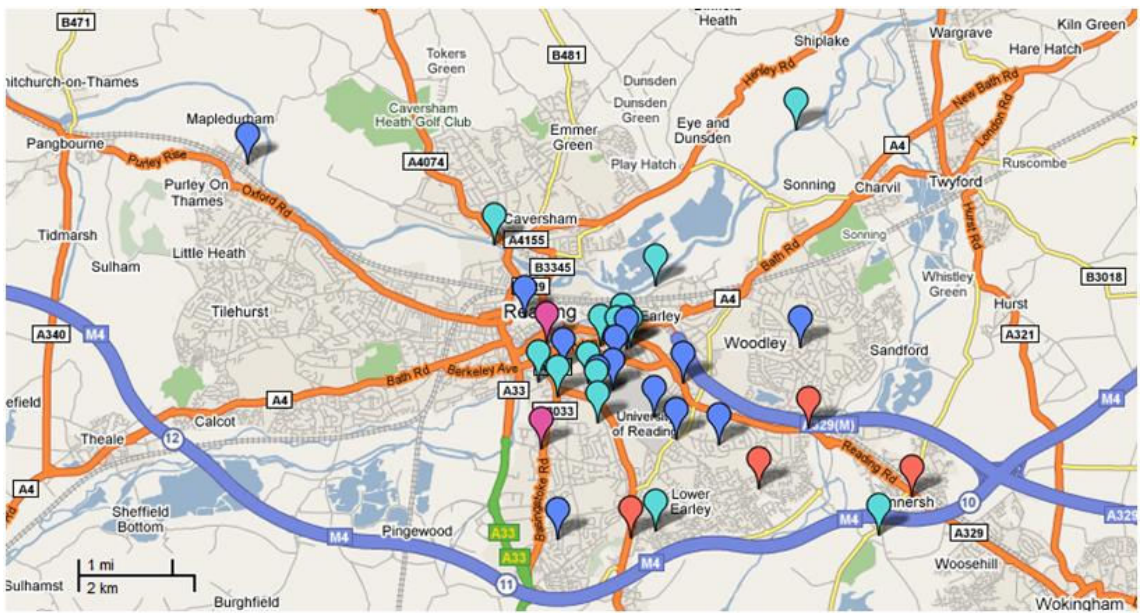

**Figure 7:** Reading (map constructed using Google Maps) on 10 January 2009. Dark blue pins are lying snow, light blue pins are falling snow, purple pins are very light falling snow, and red pins are where there was no evidence of snow at all.

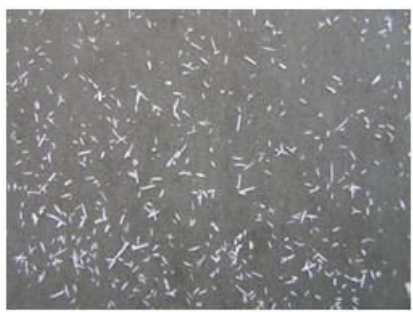

**Figure 8:** Snow grains about 5 mm long, on garden furniture in Woodley (east of Reading town centre) at 09:00 on 10 January 2009. (With permission from Mike Blackburn.)

## Closing remark

These observations presented evidence for snowfall from freezing fog during non-frontal, anticyclonic conditions. There are growing numbers of reports of these events in the UK.

## Acknowledgements

Thanks to Annette Osprey for help with the observations in Hereford; Dave Bennett, Karen Aplin, and Chris Witts for observations in Gloucester; numerous members of the University of Reading's Meteorology Department for observation around Reading (including Mike Blackburn for the photograph); UK Met Office for synoptic analyses; and CloudNet for the cloud radar images.

## Appendix

Below is the route taken to make the observations in Hereford, 31 December 2008. Times in hours UTC.

| Time | Location | Observation |
|------|----------|-------------|
| 13:20 | **A49 North** | No evidence of snow, -1.5°C outside city |
| 13:30 | **Homer Road** | No evidence of snow |
| 13:34 | **Whitecross Common** | Falling and settled snow |
| 13:35 | **Chave Court Place** | Falling and lying snow |
| 13:37 | **The Vines** | Falling and lying snow |
| 13:38 | **Grandstand Road** | Very light falling snow |
| 13:39 | **Yazor Road** | No snow |
| 13:40 | **Sherrington Drive** | No snow |
| 13:41 | **Baggallay Street** | Snow stopped falling |
| 13:42 | **Plough Lane** | Falling snow, 1°C |

| Time | Location | Observation |
|---|---|---|
| 13:43 | **Friar Street** | Lightly falling snow, 1.5°C |
| 13:44 | **Edgar Street** | Heavier snow falling |
| 13:45 | **Penhaligon Way** | Very slight falling snow |
| 13:49 | **Newtown Road** | Very very slight falling snow |
| 13:50 | **PC World** | No falling snow, snow on some cars and posts |
| 14:10 | **Widemarsh Street** | Lying snow; no falling snow to start, then light falling snow |
| 14:12 | **Blackfriars Street** | Light falling snow, lying snow, 2.5°C |
| 14:20 | **Blueschool Street** | No falling snow, snow on ground |
| 14:25 | **Cantilupe Street** | Intermittent light snowfall, lying snow |
| 14:30 | **Bartonsham Road** | Lying snow |
| 14:31 | **Outfall works road** | No snow, 3.5°C |
| 14:32 | **Hampton Park Road** | No snow |
| 14:35 | **Station approach** | Lying snow |
| 14:36 | **Rockfield Road** | No snow, 4.5°C |
| 14:38 | **Aylestone Hill** | No snow |
| 14:39 | **Venn's Lane** | No snow |
| 14:41 | **Old School Lane** | No snow |
| 14:42 | **Roman Road / Old Sch** | No snow |
| 14:44 | **Starting Gate** | No snow |
| 14:45 | **Roman Road / Kempton** | No snow |
| 14:46 | **Canon Pyon Road / Roman** | No snow |
| 14:48 | **Bakers Lane** | No snow |
| 14:49 | **Three Elms Road** | No snow |
| 14:50 | **Wordsworth Road** | Lying snow, 0.5°C |
| 14:51 | **Barrie Road** | Lying snow |
| 14:53 | **Emlyn Avenue** | Lying snow |
| 14:55 | **Westfaling Street** | Lying snow |
| 15:00 | **Breinton Road** | Lying snow, 0.5°C |
| 15:05 | **Upper Breinton** | Lying snow |
| 15:10 | **Breinton** | No snow |
| 15:15 | **Swainshill South** | No snow |
| 15:17 | **Swainshill** | No snow |
| 15:22 | **Bramley Court** | No snow |
| 15:23 | **Cotswold Drive** | Lying snow (very patchy) |
| 15:35 | **Barton Road** | Lying snow |
| 15:37 | **ASDA** | Lying snow |
| 15:39 | **Ross Road / Holme Lacy** | Lying snow |
| 15:40 | **Ross Road / Bullingham** | Lying snow |
| 15:42 | **Ross Road** | Lying snow |
| 15:44 | **Rotherwas Relief Road Roundabout** | No snow |
| 16:00 | **Coningsby Street** | Lying snow |
| 16:30 | **High Town** | Lying snow |
| 17:00 | **School** | Playground covered in snow |